\documentclass[12pt]{article}

\usepackage{color}
\usepackage{amsmath,amssymb}
\usepackage{latexsym}
\usepackage[dvips]{graphicx,psfrag}
\usepackage{cite}
\usepackage{wrapfig}
\usepackage{young}

\baselineskip=\normalbaselineskip
\renewcommand{\baselinestretch}{1.34}
\renewcommand{\thefootnote}{\fnsymbol{footnote}}

\newcommand{\vev}[1]{\left\langle #1 \right\rangle}
\newcommand{\ket}[1]{\bigl|#1\bigr>}

\newcommand{\bracket}[2]{\left.\left\langle #1\right|#2\right\rangle}
\newcommand{\maru}[1]
{{\ooalign{\hfil#1\/\hfil\crcr\raise.167ex\hbox{\mathhexbox20D}}}}

\newcommand{\bP}{\boldsymbol{P}}
\newcommand{\bQ}{\boldsymbol{Q}}

\newcommand{\bunit}{\boldsymbol{1}}

\newcommand{\bB}{\boldsymbol{B}}

\newcommand{\bL}{\boldsymbol{L}}
\newcommand{\bX}{\boldsymbol{X}}

\newcommand{\bOmega}{\boldsymbol{\Omega}}

\newcommand{\bPsi}{\boldsymbol{\Psi}}

\newcommand{\del}{\partial}

\newcommand{\eq}[1]{(\ref{#1})}

\newcommand{\nn}{\nonumber}

\DeclareMathOperator{\tr}{tr}
\DeclareMathOperator{\diag}{diag}

\allowdisplaybreaks[1]

\newcommand{\ds}{\displaystyle}

\makeatletter
\@addtoreset{equation}{section}
\@addtoreset{theorem}{section}
\@addtoreset{definition}{section}
\@addtoreset{lemma}{section}
\@addtoreset{proposition}{section}

\begin{document}
\begin{flushright}
\parbox{40mm}{%
{\tt arXiv:0902.1676} \\
February 2009}
\end{flushright}

\vfill

\begin{center}
{\Large{\bf 
Fractional supersymmetric Liouville theory \\
and the multi-cut matrix models
}}
\end{center}

\vfill

\begin{center}
{\large{Hirotaka Irie}}\footnote%
{E-mail: {\tt irie@phys.ntu.edu.tw}}   \\[2em]
Department of Physics and Center for Theoretical Sciences, \\
National Taiwan University, Taipei 10617, Taiwan, R.O.C \\
\end{center}
\vfill
\renewcommand{\thefootnote}{\arabic{footnote}}
\setcounter{footnote}{0}
\addtocounter{page}{1}

\begin{center}
{\bf abstract}
\end{center}

\begin{quote}
We point out that the non-critical version of the $k$-fractional superstring theory 
can be described by $k$-cut critical points of the matrix models.
In particular, in comparison with the spectrum structure of fractional super-Liouville theory, 
we show that $(p,q)$ minimal fractional superstring theories appear 
in the $\mathbb Z_k$-symmetry breaking 
critical points of the $k$-cut two-matrix models 
and the operator contents and string susceptibility coincide on both sides. 
By using this correspondence, 
we also propose a set of primary operators of 
the fractional superconformal ghost system 
which consistently produces the correct gravitational scaling critical exponents of 
the on-shell vertex operators. 
\end{quote}
\vfill
\renewcommand{\baselinestretch}{1.4}
\newpage

\section{Introduction}

String theory plays important roles not only as a candidate of the fundamental theory of our universe 
but also as an interesting tool to investigate other physics areas, including 
nuclear physics and condensed matter physics. 
We can also expect that there would be a possibility that 
other kinds of string theories have hidden some interesting 
connections to various regimes of physics. 

The string theory considered in this paper is {\em fractional superstring theory} of order $k$ \cite{ALT}, 
a different kind of string theory whose worldsheet gauge symmetry is 
so called {\em fractional supersymmetry} 
\cite{GKO,KMQ,Bagger,Ravanini,BL,ABL,FSsineGordon}. 
Each number $k$ gives a different kind of theory 
(the case of $k=1$ ($2$) is nothing but the usual bosonic (super) string theories). 
There are several reasons for studying these theories. One might be a phenomenological reason
to have a model of lower (or reasonable) critical dimensions less than ten \cite{AT1,ModelBuilding}.
Another can be a possibility of extending the spacetime spectrum to include different types of statistics 
\cite{CR,WSCFT,CCSS} 
(This feature might be good for applying to some systems with different kinds of statistics).
In addition, it is also interesting to deepen our understanding of the RNS superstring formulation
from the viewpoint of this generalization of worldsheet conformal field theory. 
For instance, this would be helpful to understand several structures 
among spacetime and worldsheet, like their symmetries and statistics. 

Roughly speaking, fractional superstring theory is obtained by replacing the worldsheet fermions 
in the superstring theory with the Zamolodchikov-Fateev parafermions \cite{ZF1}. 
Even though there is much progress in studying this system 
\cite{ALT,StructureConstants,BRSTCoset,NewJacobi,AT1,
ModelBuilding,LowLying,CR,Scattering,
WSCFT,CCSS,NewKac}, 
several difficulties have prevented us from revealing its whole body and structure. 
The main difficulty comes from the special feature of this fractional supersymmetry: 
spin of the current is equal to some fractional number, $(k+4)/(k+2)$ ($k=2,3,\cdots$), 
and the algebra of this current is of the non-local non-Abelian braiding type 
\cite{StructureConstants}. 
Consequently, these facts cause the complexity of the system and 
also make it also difficult to identify the appropriate ghost system. 

In this paper, we will shed new light on this string theory 
from another promising approach which is known as non-critical strings and matrix models 
\cite{Polyakov,KPZ,DDK,DHK,SeibergNotes,LianZuckerman,DOZZ,fuku-hoso,SeSh,Okuyama,Irie,
KazakovSeries,dsl,Douglas,PQop,DKK,fkn1,fkn2,fkn3,PeShe,DSS,Nappi,MultiCut,CDM,HMPN,fy1,fy2,fy3,
TT,NewHat,UniCom,MMSS,SeSh2,fis,fim,fi1}. 
We point out that the non-critical version of fractional superstring theory 
has the matrix-model dual description, known as multi-cut matrix models \cite{MultiCut}. 
This should be an important clue to the investigation 
since it enables us to extract not only perturbative information 
but also non-perturbative one (for example, the D-branes in fractional superstring theory).
The existence of a consistent fractional superstring theory is quite significant because we totally lost the reason 
why we can ignore the possibility of this string theory. Since there are infinitely many kinds 
of the fractional superstrings (which are essentially related to the classification of Kac-Moody algebra) 
\cite{GenFSST1}, 
this implication opens a broad possibility of string theory. 

In the rest of this section, we explain the basic idea of the correspondence between fractional superstring theory 
and multi-cut matrix models \cite{irieKEK}, which can be seen from the 
minimal string field formulation for the multi-cut matrix models \cite{fi1,fi3}. 
Let us first recall the two-cut-matrix-model case, 
which corresponds to type 0 superstring theory \cite{TT,NewHat,UniCom}. 

The important point of the correspondence with superstrings ($k=2$) is the fact that 
the $\mathbb Z_2$ reflection transformation of the two-cut-matrix-model eigenvalues, 
$\lambda \to -\lambda$, corresponds to the RR charge conjugation of the D-branes 
\cite{TT,NewHat,UniCom,SeSh2}. 
Thereafter, this structure was shown to be clarified 
by the minimal string field formulation \cite{fy1,fy2,fy3} of this two-cut case \cite{fi1}. 

In the minimal superstring field formulation \cite{fi1}, 
the FZZT and its anti-FZZT brane in this theory were identified with 
the two-component fermions $c_0^{(1)}(\zeta)$ and $c_0^{(2)}(\zeta)$. 
The charge conjugation is, therefore, equivalent to exchanging these two fermions:
\begin{align}
\Omega: \,\,\,c_0^{(1)}(\zeta)\quad \leftrightarrow \quad c_0^{(2)}(\zeta). 
\end{align}
One way to see the connection with the superstring theory is to consider
the bosonization field $\varphi_0^{(i)}(\zeta)$ 
($ c_0^{(i)}(\zeta)\equiv :\!e^{\varphi_0^{(i)}(\zeta)}\!:$), which turns out to be 
a string field of the FZZT-brane boundary state \cite{fy3}. 
That is, the following re-expression of the string fields, 
\begin{align}
\varphi_0^{(i)} (\zeta)\equiv \varphi^{[0]}_0(\zeta)+ (-1)^{(i-1)} \varphi^{[1]}_0(\zeta),
\end{align}
with respect to the behavior of the charge conjugation 
($\Omega:\, \varphi_0^{[\mu]}(\zeta) \to (-1)^\mu \varphi_0^{[\mu]}(\zeta)$) 
can be interpreted as the decomposition into the NSNS and RR contributions of the boundary states 
in the CFT language,
\begin{align}
\ket{B;(\zeta,i )} = \ket{B;\zeta}_{\rm NSNS}+(-1)^{i-1}\ket{B;\zeta}_{\rm RR}.
\end{align}
Note that the conservation law of the RR charge is related to the 
$\mathbb Z_2$ spin structure of the worldsheet fermion. So the existence of the $\mathbb Z_2$ charged D-branes 
is one of the indications of superstring theory. 

In the same way, the $k$-cut matrix model can be described with the $k$-component fermions $c_0^{(i)}(\zeta)$ ($i=1,2,\cdots,k$), 
and they have the following $\mathbb Z_k$-charge-conjugation property \cite{fi3}:
\begin{align}
\Omega^n: \,\,\, c_0^{(i)}(\zeta ) \quad \to \quad c_0^{(i+n)}(\zeta)\qquad (n\in \mathbb Z),
\end{align}
with $c_0^{(i+k)}(\zeta)=c_0^{(i)}(\zeta)$. 
This $\mathbb Z_k$ charge conjugation is also related to the $\mathbb Z_k$ rotation 
of the eigenvalue, $\lambda \to e^{\frac{2\pi i }{k}n}\lambda $, 
of the multi-cut matrix models (See \cite{fi1} or section 2.2). 

Therefore it is also natural to re-express the bosonization fields $\varphi_0^{(i)}(\zeta)$ ($i=1,2,\cdots,k$) as 
\begin{align}
\varphi^{(i)}_0(\zeta )\equiv \varphi_0^{[0]}(\zeta)+\omega^{i-1}\varphi_0^{[1]}(\zeta)+\cdots +\omega^{(i-1)(k-1)}\varphi_0^{[k-1]}(\zeta) \qquad (\omega\equiv e^{2\pi i/k}),
\end{align}
with respect to the behavior of the $\mathbb Z_k$ charge conjugation 
($\Omega: \varphi_0^{[\mu]}(\zeta) \to \omega^\mu\, \varphi_0^{[\mu]}(\zeta)$). 
Then the D-branes in the corresponding string theory should be expressed as 
the $\mathbb Z_k$ charged boundary states with {\em the $\mathbb Z_k$ generalized Ramond sector}:
\begin{align}
\ket{B;(\zeta,i)}= \ket{B;\zeta}_{\rm NSNS}+\omega^{i-1}\ket{B;\zeta}_{\rm RR^{[1]}}
+ \cdots + \omega^{(i-1)(k-1)} \ket{B;\zeta}_{{\rm RR}^{[k-1]}}. 
\end{align}
In this sense, it is natural to conjecture that the $k$-cut extension is 
realized by replacing the fermions in superstring theory 
with {\em parafermions} which realize the $\mathbb Z_k$ spin structure%
\footnote{The basics of Zamolodchikov-Fateev parafermion 
and its $\mathbb Z_k$ spin-structure are in Appendix A. } on worldsheet \cite{irieKEK}. 
That is, the multi-cut matrix models should correspond to the fractional superstring theory.%
\footnote{It was also pointed out in \cite{irieKEK} that at least some special (but infinitely many) 
Kac tables of the unitary $(p,q)=(p,p+k)$ minimal fractional SCFT 
of the GKO coset construction \cite{GKO} coincides with that of the differential operator system of 
the $k$-component KP hierarchy of $k$-cut matrix models.}
In particular, we will show that the simplest class of $k$-fractional superstring theory (i.e. 
$(p,q)$ minimal $k$-fractional superstring theory) has the same spectrum structure and string susceptibility 
as those in the $k$-cut two-matrix model.

The organization of this paper is following: 
In section 2, we first give the definition and meaning of the multi-cut two-matrix models 
and then discuss the spectrum in the critical points.  
In section 3, we study fractional super-Liouville theory, 
especially the matching of the operator contents and string susceptibility in both sides.
After that, we discuss the gravitational scaling exponents and the ghost primary fields 
in the fractional super-Liouville theory. 
Section 4 is devoted to conclusion and discussion. 
In the appendix A, we summarize the basics of parafermion and 
introduce some notations and terminology.

\section{The multi-cut matrix models}
\subsection{The meaning of the multi-cut matrix integrals}

Here we first note the meaning and definition of the multi-cut matrix model 
and then show its operator contents in next subsection. 

\begin{figure}[htbp]
\begin{center}
\includegraphics[scale=0.9]{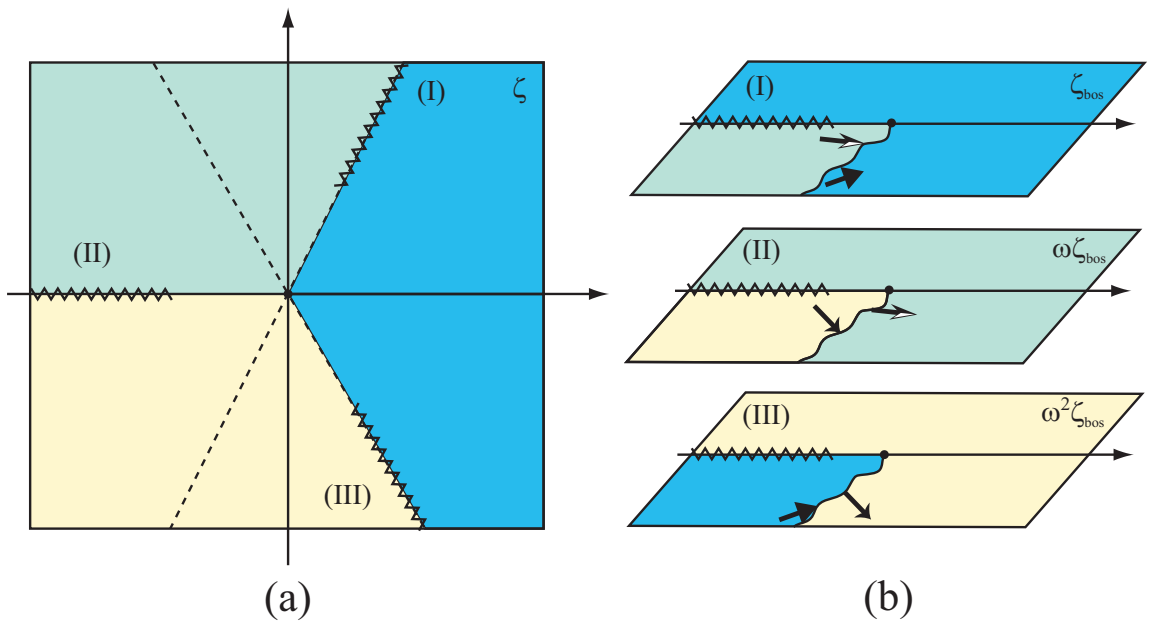}
\end{center}
\caption{\footnotesize The case of $k=3$: 
(a) The multi-cut geometry of the matrix models. The cuts form radial pattern. 
This means that there are $\mathbb Z_k$ mirror images of the spacetime. 
(b) The spacetime (Liouville-theory) geometry in terms of $\zeta_{bos}=\zeta^k$. 
There are $k$ sheets (spacetime), and there is only one cut along the real axis on each sheet.  }
\label{cuts}
\end{figure}

The meaning of ``multi-cut'' in the multi-cut matrix models was first proposed by 
C. Crnkovid and G. Moore \cite{MultiCut}. Since the double scaling limit means that 
we only focus on the vicinity of the critical points, the general multi-cut configuration 
should have the cuts which run in a radial pattern (See Figure. \ref{cuts} (a)). 
This can be understood as ``orbifold matrix models,'' 
which mean that the hermit matrix $H$ is replaced 
with some matrix $\Phi$ such that $\Phi^k=H$ is hermit:%
\footnote{Note that this configuration is different from the configuration where all the multi (more than two) cuts 
run only along the real axis in the planer limit of the matrix models, 
which is also called ``multi-cut matrix models'' in the literature.} 
\begin{align}
\int DH\, e^{-N\tr V(H)}\quad  \to \quad  \int D\Phi \,e^{-N\tr \tilde V(\Phi)}.
\end{align}

This orbifolding picture is also justified by the recent progress in the two-cut matrix models \cite{UniCom,SeSh2,fi1} which came after the original investigations \cite{PeShe,DSS,Nappi,CDM,HMPN}. 
In particular with the spacetime interpretation of the boundary cosmological constant 
$\zeta_{bos} = e^{-b(\phi+iX)}$ \cite{fy3,MMSS,SeSh2}, 
it was argued by \cite{fi1} that 
the multi-cut matrix models represent ``orbifolding of spacetime'' 
and that the multi-cut geometry of the multi-cut matrix models which is probed by cosmological constant $\zeta$ 
is the $k$-th root of the original spacetime, $\zeta_{bos}\, (= \zeta^k)$, 
on which there is only one cut along the real axis on each sheet (Figure \ref{cuts} (b)).
We can also consider some condensation of twisted modes in spacetime.%
\footnote{The R-R sector in the two-cut case. }
This is nothing but the consideration of the matrix-model potentials which breaks the $\mathbb Z_k$ symmetry 
\cite{Nappi}. 

\

The definition of multi-cut matrix models is usually given 
in the orthogonal polynomial method of the diagonalized matrix models 
since the essential information (the integrable hierarchy and string equations) 
comes from the recursive relations of the polynomials \cite{MultiCut}. 

How to define multi-cut ``multi-matrix'' models was given in \cite{fi1} as is shown below:
The orthogonal polynomial system of the two-matrix models is as usual except for the 
contour $\mathcal C_k \times \mathcal C_k \subset \mathbb C^2$ of the integration of $(x,y)$:
\begin{align}
\bracket{\alpha_n}{\beta_m}\equiv 
\int_{\mathcal C_k\times \mathcal C_k} d x dy\, 
e^{-Nw(x,y)}\,\alpha_n(x)\,\beta_m(y) = \delta_{n,m},  \label{Orth}
\end{align}
with the orthonormal polynomials, 
\begin{align}
\alpha_n(x)=\frac{1}{\sqrt h_n} x^n + \cdots, \quad 
\beta_n(y) =\frac{1}{\sqrt h_n} y^n + \cdots,
\end{align}
and the potential $w(x,y)=V_1(x)+V_2(y)-cxy$. 
The contour $\mathcal C_k \times \mathcal C_k\subset \mathbb C^2$ 
of the integration is 
\begin{align}
\mathcal C_k \equiv \bigl\{\omega^n z \in \mathbb C\, ; z \in \mathbb R, \, n\in \mathbb Z \bigr\},
\end{align}
with the parameter $\omega=e^{2\pi i/k}$. 
The proposal of how to construct the multi-cut matrix model \cite{fi1} is: (i) to consider 
the following simultaneous $Z_k$ transformation of eigenvalues $(x,y)$: 
\begin{align}
\Omega^n:\, (x,y) \mapsto (\omega^n x,\omega^{-n} y) 
\qquad (n=0,1,2,\cdots,k-1), 
\end{align}
which should be respected in the $k$-cut models, 
and (ii) to focus on the critical behavior in the vicinity of the fixed point $(x,y)=(0,0)$. 
In the special case of two-cut two-matrix models, 
this system correctly reproduces the results of super-Liouville theory, 
at least with some scaling ansatz \cite{fi1}. 

One may wonder how we can realize this multi-cut geometry (especially $k\geq 3$) 
in terms of ``matrix-model integral''. 
That is, how to specify the matrix $\Phi$ and their measure $D \Phi$ in its matrix model.
It is easy to construct the matrix integral 
for multi-cut multi- (two- in practice) matrix models, and 
fortunately, it is the case where 
the matrix model should have the correspondence with 
minimal fractional superstrings. 
When we try to construct ``multi-cut matrix quantum mechanics'', 
we may find it is not so straightforward. 
This should be, however, just a technical problem because 
2D fractional superstring theory would be consistently defined, 
at least from the the worldsheet point of views.%
\footnote{
Only when $k\leq 2$, the two-matrix models can include the one-matrix models 
as the special case of the system because the Gaussian potential $V(y)=y^2$ which is necessary 
for the reduction into one-matrix model is forbidden in the higher $\mathbb Z_k$ symmetric case. }

The integral is given as follows:
The important requirement here is that the measure of the inner product \eq{Orth} should be invariant 
under this $\mathbb Z_k$ transformation, 
even though the matrix-model potentials $V_1(x)$ and $V_2(y)$ is not necessarily symmetric.  
Note that, with these requirements, 
the orthonormal conditions \eq{Orth} are consistent under this $\mathbb Z_k$ transformation. 
From this orthogonal polynomial system, 
we can define the meaningful matrix integral, 
\begin{align}
\mathcal Z_{MM}  = \int_{\mathcal C_{M_k}\times \mathcal C_{M_k}} dXdY\, e^{N\tr w(X,Y)}. 
\end{align}
The measure is defined by the following metric:
\begin{align}
ds_X^2=\tr\bigl[(dX)^2\bigr],\qquad ds_Y^2=\tr\bigl[(dY)^2\bigr],
\end{align}
and then the measure $dXdY$ is invariant under the $\mathbb Z_k$ transformation,%
\footnote{We cannot use this measure when the number of the matrices in the model 
is odd, which includes one-matrix models and matrix quantum mechanics. }
\begin{align}
(X,Y)\to (\omega^n X,\omega^{-n}Y).
\end{align}
The contour $\mathcal C_{M_k} \subset gl(N,\mathbb C)$ of the matrix integral should be 
\begin{align}
\mathcal C_{M_k} \equiv 
\{U\diag (x_1,x_2,\cdots,x_N)\,U^\dagger\, 
;\, U\in U(N), x_i\in \mathcal C_k \} \subset gl(N,\mathbb C).
\end{align}
This means that the matrix $X$ (and $Y$) of this model should be an $N \times N$ normal matrix $X$
such that $X^{\hat k}=H$ is an hermit matrix.%
\footnote{This kind of matrix contour integral is also observed in the supermatrix description of the type 0 superstrings \cite{fi2}.} 
Here we define, 
\begin{align}
\tilde k=
\left\{
\begin{array}{ll}
k & (k \in 2\mathbb Z+1) \cr
k/2 & (k\in 2 \mathbb Z). 
\end{array}
\right.
\end{align}
Every odd $k$-cut model (i.e. $k$ is odd) has the same contour integral 
as even $2k$-cut models have. 
They are, however, physically different systems because their Liouville directions 
$e^{-b\phi}={\rm Re}\,(\zeta_{bos})={\rm Re}\,(\zeta^k)$ are different. 

\subsection{The spectrum of the multi-cut two-matrix models}

In general, the matrix model possesses the integrable structure of KP hierarchy and 
the string field formulation \cite{Douglas,PQop,fkn1,fkn2,fkn3,fy1,fy2,fy3}.
The two-cut matrix models also have the two-component extension of the KP hiererchy \cite{fi1}. 
These structures are really useful to see the physical picture and dynamics of the string theory \cite{fy1,fy2,fy3,fis,fim}.
This formulation is also powerful even for the case of multi-cut matrix models
and enable us to see various information of the matrix models.  

From the case of two-cut matrix models \cite{fi1}, it is obvious that 
the integrable hierarchy of the $k$-cut matrix model is given by 
the $k$-component KP hierarchy, and 
the extension to the case of the $k$-component KP hierarchy is 
essentially the same as the extension to the two-component KP case \cite{fi3}.%
\footnote{The author would like to thank Prof. Masafumi Fukuma for the various useful discussion and sharing insights into this matter.}
So we just briefly mention how the $k$-component KP hierarchy appears in the $k$-cut two-matrix models 
to extract information of the spectrum. 
See \cite{fi1} for more detail discussion, and \cite{KaLe} for the $k$-component KP hierarchy. 

The integrable structure comes from the recursive relations of the orthonormal polynomials, 
\begin{align}
x\, \alpha_n(x)= \hat A_x(n, e^{\del_n}) \cdot \alpha_n(x),& \qquad 
N^{-1}\frac{\del}{\del x}\, \alpha_n(x)= \hat B_x(n,e^{\del_n}) \cdot \alpha_n(x), \nn\\
y\, \beta_n(y)= \hat A_y(n, e^{\del_n}) \cdot \beta_n(y),& \qquad 
N^{-1}\frac{\del}{\del y}\, \beta_n(y)= \hat B_y(n,e^{\del_n}) \cdot \beta_n(y),
\end{align}
with the canonical commutation relations, 
$[\hat A_x,\hat B_x]=[\hat A_y,\hat B_y]=N^{-1}$. 
For the simplicity, we first choose the $\mathbb Z_k$ symmetric potentials 
$V_1(\omega x)=V_1(x)$ and $V_2(\omega y) = V_2(y)$, then we can see the $\mathbb Z_k$ 
transformation property of the orthonormal polynomials:
\begin{align}
\alpha_n(\omega x)=\omega^n \alpha_n(x), 
\qquad \beta_n(\omega y)=\omega^n \beta_n(y).
\end{align}
If we have critical points near the origin $(x,y)=(0,0)$, 
then there should be the $k$ kinds of scaling orthonormal functions $\{\hat \Psi_n(t;\zeta)\}_{n=1}^k$ 
which satisfy the same $\mathbb Z_k$ transformation property: 
\begin{align}
\hat \Psi_n(t;\omega \zeta)= \omega^n \,\hat \Psi_n(t;\zeta), 
\end{align}
or equivalently, the scaling functions $\{\Psi_i(t;\zeta)\}_{i=1}^k$ which satisfy
\begin{align}
\Psi_i(t;\omega \zeta)=\Psi_{i-1}(t;\zeta),\qquad 
\Psi_i(t;\zeta)=\sum_{n=0}^{k-1} \omega^{-ni}\, \hat \Psi_n (t;\zeta).
\end{align}
Furthermore, if we can find the special scaling limits, 
\begin{align}
x = \zeta\, a^{\frac{\hat p}{2}} ,\qquad N^{-1} = g \, a^{\frac{\hat p +\hat q}{2}}, 
\qquad \frac{n}{N}=1+ t\, a^{\frac{\hat p+\hat q-1}{2}},
\qquad \del_n=(g \, \del_t)\, a^{\frac{1}{2}},
\end{align}
which make the operator $\hat A$ and $\hat B$ the differential operators $(\bP,\bQ)$ 
in $\del\, (\equiv \del_t)$ of $(\hat p,\hat q)$ order,%
\footnote{We should note that there should be several nontrivial changes of 
the basis of orthonormal polynomials which depend on the index $n$ of $\alpha_n(x)$
to get the smooth scaling functions $\Psi_i(t;\zeta)$. For example, in the two-matrix model case, 
we need the change of the overall sign of the functions: $\alpha_n(x) \to (-1)^{[n/4]}\alpha_n(x)$ 
($[*]$ is the Gauss symbol).
Here we just assume the existence of scaling functions, 
which should be checked by some direct evaluation of critical points. }
\begin{align}
a^{-\hat p/2}\hat A_x(n,e^{\del_n}) \quad \leadsto \quad 
\bP(t,\del) = U_0^{(P)} \del^{\hat p} + U_1^{(P)}(t) \del^{\hat p-1} + \cdots + U_{\hat p}^{(P)}(t), \nn\\
a^{-\hat q/2}\hat B_x(n,e^{\del_n}) \quad  \leadsto \quad
\bQ(t,\del)= U_0^{(Q)} \del^{\hat q} + U_1^{(Q)}(t) \del^{\hat q-1} + \cdots + U_{\hat q}^{(Q)}(t),
\end{align}
then the operator $(\bP,\bQ)$ should be $k \times k$ matrix valued differential operators. 
At this level, we can consider $\mathbb Z_k$ breaking potential even such a critical points \cite{Nappi, CDM,HMPN}.
With these operators, we can see that the scaling recursive relations turn out to be the differential equations, 
\begin{align}
\zeta \,\Psi(t; \zeta ) = \bP(t,\del) \, \Psi(t;\zeta),\qquad 
g \frac{\del}{\del \zeta} \Psi(t;\zeta) = \bQ(t,\del) \, \Psi(t;\zeta), \label{Baker}
\end{align}
with $\Psi(t;\zeta)\equiv {}^t(\Psi_1,\Psi_2,\cdots,\Psi_k)$, 
which are related to the Baker-Akhiezer function 
with the Douglas equation $[\bP,\bQ]=g \bunit$. 
Since the operators are $k\times k$ matrix differential operators, 
we have $k$ independent solutions $\{\Psi^{(i)}(t,\zeta)\}_{i=1}^k$, 
\begin{align}
\zeta_i \,\Psi^{(i)}(t; \zeta ) = \bP(t,\del) \, \Psi^{(i)}(t;\zeta),\qquad 
g \frac{\del}{\del \zeta_i} \Psi^{(i)}(t;\zeta) = \bQ(t,\del) \, \Psi^{(i)}(t;\zeta).
\end{align}
Note that we take $\Psi^{(1)}(t,\zeta)=\Psi(t,\zeta)$ is a special solution $\zeta_1=\zeta$ 
which satisfy \eq{Baker}. 
In the superstring case $(k=2)$, this another solution $\Psi^{(2)}$ is related to 
the anti-FZZT-branes which have different charges from the original $\Psi^{(1)}$. 
In particular, each solution $\Psi^{(i)}(t;\zeta)$ corresponds to each free fermion 
of the $k$-component KP hierarchy, $c_0^{(i)}(\zeta)$ (See \cite{fi1} for how to relate). 
Thus it is natural to say that $\Psi^{(i)}$ (or $c_0^{(i)}(\zeta)$) corresponds to 
$\mathbb Z_k$ charged D-branes.

In general, the matrices $U_0^{(P)}$ and $U_0^{(Q)}$ commute with each other, thus they can be chosen to be 
diagonal matrices. Notice that the diagonal elements of $U_0^{(P)}$ are 
directly related to the eigenvalues $\zeta_i$ of 
each solution $\Psi^{(i)}(t,\zeta)$, and that we can freely change 
the relations among $\{\zeta_i\}_{i=1}^k$ and $\zeta$.%
\footnote{It is because this is just a redefinition of the boundary cosmological constants $\{\zeta_i\}$ 
of each FZZT brane. See \cite{fi1}.}
That is, we can freely (without loss of generality) chose the matrix $U_0^{(P)}$ as%
\footnote{The reason why this form is canonical can be easily seen as follows: For example, consider the $k=3$ case. 
If we have the $Z_k$ symmetric system, then the recursive relation of $\zeta \times$ (resp. $\del_\zeta\times$) 
\eq{Baker} should be a map from $\hat \Psi_n(t;\zeta)$ to $\hat \Psi_{n+1}(t;\zeta)$ 
(resp. from $\hat \Psi_n(t;\zeta)$ to $\hat \Psi_{n-1}(t;\zeta)$). 
Thus the differential operator $\bP$ and $\bQ$ should start from the shift matrices:
\begin{align}
\bP=
\begin{pmatrix}
    & 1 &  \cr
    &    & 1 \cr
1  &    &  
\end{pmatrix}
\del^{\hat p} + \cdots,\qquad 
\bQ=
\begin{pmatrix}
   &    & 1 \cr
1 &    &    \cr
   &  1  &  
\end{pmatrix}
\del^{\hat q}+\cdots, \label{Foot11}
\end{align}
the diagonalization of which gives \eq{Up}. \label{ZkSym}
}
\begin{align}
U_0^{(P)}= \Omega \equiv 
\begin{pmatrix}
1         &              &             &                        \cr
           & \omega &            &                          \cr 
            &               & \ddots &                        \cr
           &               &            &                \omega^{k-1} 
\end{pmatrix}, \label{Up}
\end{align}
which gives 
\begin{align}
\bP(t,\del) \,\bPsi(t,\zeta) = \bPsi(t,\zeta)\, Z,\qquad 
\bQ(t,\del)\, \bPsi(t,\zeta) = g \,\bPsi(t,\zeta)\,\overleftarrow{\frac{\del}{ \del Z}},
\end{align}
with 
\begin{align}
Z\equiv 
\begin{pmatrix}
\zeta        &              &             &                        \cr
           & \omega\, \zeta  &            &                          \cr 
            &               & \ddots  &                        \cr
           &               &            &                \omega^{k-1}\, \zeta 
\end{pmatrix},\qquad 
\overleftarrow{\frac{\del}{\del Z}}\equiv 
\begin{pmatrix}
\overleftarrow{\del_\zeta }       &              &             &                        \cr
           &  \overleftarrow{\del_\zeta}\,\omega^{-1}  &            &                          \cr 
            &               & \ddots  &                        \cr
           &               &            &                 \overleftarrow{\del_\zeta }\,\omega^{1-k}
\end{pmatrix},
\end{align}
and $\bPsi(t; \zeta)=(\Psi^{(1)}(t;\zeta),\Psi^{(2)}(t;\zeta),\cdots,\Psi^{(k)}(t;\zeta))$, 
an $N\times N$ matrix-valued function.%
\footnote{From these expression, we can easily see that, if we choose $\Psi^{(2)}$ as the starting point of the 
function \eq{Baker}, we need to change the boundary cosmological constant as $\zeta \to \omega^{-1}\zeta$. 
Then the role of each function $\Psi^{(i)}$ shifts as follows: $\Psi^{(i)} \to \Psi^{(i-1)}$. 
So there is no priory reason to choose $\Psi^{(1)}$ and 
this $\mathbb Z_k$ nature gives the ``$\mathbb Z_k$ charge'' of D-branes. 
(This is the same reason we can also assign a positive electric charge to electron.)}

With this differential operator $\bP(x,\del)$, 
we can construct the Lax operators of the $k$-component KP hierarchy, $\bL$ and $\bOmega$, 
\begin{align}
\bL=\del +\sum_{n=0}^\infty U_n(t)\, \del^{-n}, \qquad 
\bOmega= \Omega + \sum_{n=1}^\infty H_n(t) \,\del^{-n}. 
\end{align}
which satisfy $[\bL,\bP]=[\bOmega^n,\bP]=0$ $(n =1,2,\cdots,k)$ and $\bP =(\bOmega \bL^{\hat p})$. 
In this sense, we can have the operators $\bQ$ in terms of $\bL$ and $\bOmega$ \cite{Krichever,fkn3}:
\begin{align}
\bQ(b;\del) = 
\sum_{n=1}^{\hat p+\hat q}\sum_{\mu=0}^{k-1} 
\frac{nb^{[\mu]}_{n}}{\hat p } (\bOmega^{\mu-1} \bL^{n-\hat p})_+
\end{align}
with the general background $b=\{b_n^{[\mu]}\}$. 
From these operators $(\bP,\bQ)$, we can obtain the spectrum of the k-cut matrix models 
as the KP flow of this system \cite{n-comp.KP}:
\begin{align}
\bB^{[\mu]}_n \equiv (\Omega^\mu \bL^n)_+, \qquad 
g \frac{\del}{\del t_n^{[\mu]}} \bX(t,\del) = [\bB_n^{[\mu]}, \bX(t,\del)],
\end{align}
where $\bX(t,\del)=\bP(t,\del),\bQ(t,\del),\bL(t,\del)$ and $\bOmega(t,\del)$, and we abbreviate 
the KP times $\{ t^{[\mu]}_n \}_{n,\mu}$ as $t = \{ t^{[\mu]}_n \}_{n,\mu}$. 
The original $t$ should be $t=t_1^{[0]}$. The special feature of the multi-component KP hierarchy is 
the $\mathbb Z_k$ indices $\mu$ of $t^{[\mu]}_n$ which indicate the behavior 
of the $\mathbb Z_k$ charge conjugation \cite{fi1}. 

Therefore, the gravitational scaling exponents of this spectrum are 
\begin{align}
\bB_n^{[\mu]} \sim a^{-n/2},
\end{align}
with the lattice spacing $a$. The scaling of the most relevant operators $t^{[\mu]}_1$ and 
the ``should-be'' cosmological constant $\rho \equiv t^{[0]}_{\hat q-\hat p}$ are 
\begin{align}
t^{[\mu]}_1 \sim a^{-\frac{\hat p+\hat q -1}{2}},\qquad 
\rho\equiv t^{[0]}_{\hat q-\hat p} \sim a^{-\hat p},
\end{align}
and the string susceptibility of the cosmological constant $\gamma_{str}^{(Mat)}$ is 
\begin{align}
\mathcal F_0(\rho)\sim \rho^{2-\gamma_{str}^{(Mat)}},\qquad 
\gamma_{str}^{(Mat)}=1-\frac{\hat q}{\hat p}, \label{StrSusc}
\end{align}
where $\mathcal F_0(\rho)$ is the scaling genus-zero free energy of the matrix model
(or the partition function of the genus-zero worldsheet random surfaces), and 
the gravitational scaling dimension of each operator is given as 
\begin{align}
\vev{\alpha_{n_1}^{[\mu_1]} \cdots \alpha_{n_l}^{[\mu_l]}}_{\rm c} 
\equiv \frac{\del^l \ln \mathcal F_0(t;\rho)}
                {\del t_{n_1}^{[\mu_1]}\cdots \del t_{n_l}^{[\mu_l]}}
                \Bigr|_{t=0,\rho\neq 0}
\sim \rho^{\sum_{i=1}^l\frac{n_i-(\hat p+\hat q)}{2\hat p}}. \label{GravScale}
\end{align}
Also as as important information of the differential operators $\bP$ and $\bQ$, 
the special KP flows of $\bP^n$ and $\bQ^n$ are trivial flows of the system,
which should be unphysical-state propagations of the corresponding string theory \cite{fkn3}. 
In particular, since we have $\bP=(\bOmega \bL^{\hat p})$ as the canonical choice, 
the following operators
\begin{align}
\bB_{m\hat p}^{[m]} = \bP^m =(\bOmega^{m}\bL^{m\hat p}),
\end{align}
are trivial flows of the system, and especially if we consider $\mathbb Z_k$ symmetric critical points, 
we should have%
\footnote{Diagonalization of of $\bQ$ in \eq{Foot11} gives \eq{Qsym}. See footnote \ref{ZkSym}. }
\begin{align}
\bQ=\sum_{n=1}^{\hat q+\hat p} \frac{nb^{[0]}_n}{\hat p} (\bOmega^{-1}\bL^n) + \cdots, \label{Qsym}
\end{align}
and some flows of $\bB_{n\hat q}^{[-n]}+ \cdots $ should also be trivial. 

Considering these information as inputs of the multi-cut matrix models, 
in next section, we will study the fractional super-Liouville theory and its comparison to the multi-cut matrix models. 

\section{Fractional super-Liouville theory}

\subsection{The Kac table of $(p,q)$ minimal fractional SCFT}
We first discuss comparison to the Kac table of the minimal fractional superconformal models, 
which can be derived from the generalized Feigin-Fuchs construction developed in \cite{KMQ,BRSTCoset}. 
The basic properties of parafermion and its terminology are summarized in Appendix A. 
The system is described with one free boson $X(z)$ with a background charge $\tilde Q$
and the Zamolodchikov-Fateev parafermion system $Z_k$.
The action can be written as
\begin{align}
S_{Matt} = \frac{1}{2\pi k}\int d^2z \Bigl(\del X\bar \del X +i\tilde Q\sqrt{g}\, R X\Bigr) 
+ S_{{Z}_k}(\psi^{M},\tilde \psi^M), 
\end{align}
where $S_{Z_k}(\psi^M,\tilde \psi^M)$ is the action of the parafermion $\psi^M(z)$. 
Note that here we use the $\alpha'=k$ convention. 
The superscript $M$ means that this is the FSUSY partner of matter bosonic field $X(z)$. 
The energy momentum tensor $T^M(z)$ and the fractional supercharge $G^M(z)$ 
is given as \cite{ALT, FSsineGordon}%
\footnote{We used the same normalization for the primaries of the parafermion, $\epsilon$ and $\eta$, as in 
\cite{ZF1, ALT,StructureConstants}. }
\begin{align}
T^{M}(z)&= -\frac{1}{k}\bigl(\del X(z)\bigr)^2 + i\frac{\tilde Q}{k}\del^2 X(z) + T_{Z_k}^M(z), \\
G^M(z) &=\Bigl(\del X(z) -i \frac{(k+2)\tilde Q}{4} \del \Bigr)\epsilon^M(z) 
- i \frac{kQ}{k+4}\eta^M(z). \label{SuperChargeMatt}
\end{align}
with the background charges $\tilde Q$ and $Q$, and the central charge $c_M$,
\begin{align}
\tilde Q= b-\frac{1}{b}, \qquad 
Q=b+\frac{1}{b}, \qquad 
\hat c_{M}&\equiv \frac{k+2}{3k}c_M 
= 1-2\frac{(k+2)}{k^2}\tilde Q^2. 
\end{align}
The operator $\epsilon^M(z)$ is the first energy operator of matter parafermion 
and $\eta^M(z)$ is its first descendent in the sense of $W_k$-algebra \cite{KMQ}. 
The basic fractional super-primary operators%
\footnote{This means that $G_r\cdot  \mathcal O(z)=0,\,\,  (r>0)$, for the modes of 
the fractional supercurrent $G(z)$. } 
in this Feigin-Fuchs construction are given by
\begin{align}
\mathcal O_p^{[\lambda]}(z) = \sigma_{\lambda}(z)\, :\!e^{i\frac{2}{k}pX(z)}\! :, \qquad 
\bigl(\mathcal O_p^{[\lambda]}\bigr)^\dagger(z) = \mathcal O_{\tilde Q-p}^{[k-\lambda]}(z),
\end{align}
with the dimension 
\begin{align}
\Delta(\mathcal O_p^{[\lambda]})=\Delta(\sigma_\lambda)-\frac{1}{k}p(\tilde Q-p), \qquad 
\Delta(\sigma_\lambda)=\frac{\lambda(k-\lambda)}{2k(k+2)}.
\end{align}
Here $\sigma_\lambda(z)$ is the spin field (See Appendix A). 
The Verma module of Feigin-Fuchs construction is generated by this primary: 
\begin{align}
\mathcal V_{p,\lambda}\equiv \Bigl[\mathcal O_p^{[\lambda]}(z)\Bigr]_{\epsilon^M,X} = 
\Bigl[\mathcal O_p^{[\lambda]}(z)\Bigr]_{G^M}.
\end{align}
The screening charge is defined with the basic parafermion $\psi(z)$ and $\psi^\dagger(z)$ as \cite{KMQ}
\begin{align}
\mathcal Q_\pm \equiv \oint dz\, S_\pm(z),\qquad 
S_+ =\psi(z) :\!e^{2ibX(z)/k}\! :,\qquad 
S_- =\psi^\dagger(z) :\!e^{-2iX(z)/kb}\! :, \label{Screening}
\end{align}
with $\Delta(S_\pm)=1$ and $\bigl[\mathcal Q_\pm,G^{M}(z)\bigr]=0$ \cite{KMQ}. 
The degenerate fields of the fractional superconformal field theory are given as \cite{BRSTCoset}
\begin{align}
\mathcal O_{r,s}(z)\equiv \mathcal O_{p_{r,s}}^{[r-s]}(z),
\qquad p_{r,s}=-\frac{1}{2}\Bigl((1-r)\,b^{-1}-(1-s)\,b\Bigr).
\end{align}
Here we always write $[r-s]$ to indicate that 
$r-s$ should be understood as modulo $k$. 
The first null state in the module of $\mathcal V_{r,s}$ appears at the level of 
$N_{r,s}$ \cite{BRSTCoset},
\begin{align}
N_{r,s}
=&\frac{rs}{k}+\Delta(\sigma_{[r+s]})-\Delta(\sigma_{[r-s]}), 
\end{align}
as 
\begin{align}
\chi_{r,s}(z)=\Bigl(\mathcal Q_+^{(s)}\cdot \mathcal O_{-r,s} (z)\Bigr)^\dagger 
= \Bigl(\mathcal Q_-^{(r)}\cdot \mathcal O_{r,-s} (z)\Bigr)^\dagger \in \mathcal V_{r,s}.
\end{align}

The minimal model means that 
some finite set of degenerate primary operators are closed in the OPE algebra, and it is always realized 
if the parameter $b$ is related to a rational number \cite{BPZ},
\begin{align}
b=\sqrt{\frac{\hat p}{\hat q}}. 
\end{align}
Here we assume that $(\hat p,\hat q)$ are coprime integers and $\hat q>\hat p>0$.%
\footnote{Note that these indices are related to the exponent of differential operator in 
$k$-component KP hierarchy, so we used the same notation. One may want to write the 
labeling of minimal models by $(\hat p,\hat q; k)$ like in \cite{MultiCut}. }
Then the dimension of the primary operator is written as 
\begin{align}
\Delta(\mathcal O_{(r,s)}) = \frac{(\hat qr-\hat ps)^2-(\hat q-\hat p)^2}{4k\hat p\hat q} 
+  \Delta(\sigma_{[s-r]}). 
\end{align}
The usual conformal labeling of $(p,q)$ is the range of the indices $(r,s)$: 
\begin{align}
1\leq r \leq p-1,\qquad 1\leq s\leq q-1,
\end{align}
or we can also say that the minimal numbers $(p,q)$ which satisfy 
\begin{align}
\mathcal O_{r+lp,s+lq}(z)=\mathcal O_{r,s}(z)\qquad (l\in \mathbb Z),
\end{align}
in the Feigin-Fuchs terminology.
Because of this property, there appears additional basic null state:
\begin{align}
\tilde \chi_{r,s}(z)=\mathcal Q_+^{(q-s)}\cdot \mathcal O_{-(p-r),(q-s)} (z)
= \mathcal Q_-^{(p-r)}\cdot \mathcal O_{(p-r),-(q-s)} (z) \in \mathcal V_{r,s},
\end{align}
at the level of 
\begin{align}
\tilde N_{r,s}
=\frac{(p-r)(q-s)}{k}+\Delta(\sigma_{[q+p-r-s]})-\Delta(\sigma_{[r-s]}),
\end{align}
and the labeling $(p,q)$ appear in those formulas. 
One can easily see that this conformal labeling $(p,q)$ is given as 
\begin{align}
(p,q) \equiv (\hat k \hat p,\hat k \hat q), \label{PQ}
\end{align}
where $\hat k$ is defined by $k=\hat k \cdot d_{\hat q-\hat p}$ so that $d_{\hat q-\hat p}$ is
the largest common divisor among the integers $k$ and $\hat q-\hat p$.%
\footnote{This means that $q-p\equiv 0$ $\hbox{mod.}\,k$. 
We should note that another coprime labeling of the minimal model, $(A,B)$, which was introduced 
in \cite{BRSTCoset} to 
describe the fractional level coset model $SU(2)_{k}\otimes SU(2)_{A/B-2}/SU(2)_{k+A/B-2}$ 
of $(p,q)=(A,A+kB)$, is equivalent to our labeling \eq{PQ}.}
Note that if we consider the special case where $k$ is a prime number, 
then we have two kinds of models: 
\begin{align}
(p,q) = 
\left\{
 \begin{array}{ll}
  (k \hat p,k \hat q) & :\mbox {$\hat q-\hat p \not\equiv 0$ \ (mod. $k$)}\\
 (\hat p,\hat q) &: \mbox {$\hat q-\hat p \equiv 0$ \ (mod. $k$)}. 
 \end{array}
\right.
\end{align}
If we take $k=2$, for each case they are called even and odd models respectively.
In this sense, this argument is consistent with the condition argued in \cite{DSZ}. 
Therefore the above $(p,q)$ indices are the natural generalization 
of the constraints in minimal superconformal field theory $(k=2)$, and there are several 
distinct kinds of minimal models in each $k$-th fractional superconformal field theory. 
The number of the kinds is give by the number of divisors of $k$ plus one. 
Here we show some examples of the Kac table 
(Table \ref{Kac1} and Table \ref{Kac2} in the matrix-model language). 

\begin{table}
\begin{center}
\begin{tabular}{|c|c|c|c|c||c|}
\hline
   &  &  &  &  & $1$ \\
 \hline \hline
   &  &  &  &  & $\Omega L^2$ \\
\hline
   &  &  &  & $\Omega L$ & $\Omega^2 L^4$ \\
\hline
   &  &  & $\Omega$ & $\Omega^2 L^3$ & $ L^6$ \\
\hline
   &  &  & $\Omega^2L^2$ & $L^5$ & $ \Omega L^8$ \\
\hline
   &  &$\Omega^2L$  & $L^4$ & $\Omega L^7$ & $ \Omega^2 L^{10}$ \\
\hline
   &  $\Omega^2$ &$L^3$  & $\Omega L^6$ & $\Omega^2 L^9$ & $ L^{12}$ \\
\hline
   &  $L^2$ &$\Omega L^5$  & $\Omega^2 L^8$ & $L^{11}$ & $\Omega L^{14}$ \\
\hline
  $L$ &  $\Omega L^4$ &$\Omega^2 L^7$  & $L^{10}$ & $\Omega L^{13}$ & $\Omega^2 L^{16}$ \\
\hline \hline
  $\Omega L^3$ &  $\Omega^2 L^6$ &$L^9$  & $\Omega L^{12}$ & $\Omega^2 L^{15}$ & $L^{18}$ \\
\hline
\end{tabular}
\end{center}
\caption{\footnotesize Kac table for $(\hat p,\hat q; k)=(2,3;3)$. Thus $(p,q)=(6,9)$ and 
$d_{\hat q-\hat p}=1$. The relation is $\mathcal O_{r,s}=(\bOmega^{r-s}\bL^{\hat qr-\hat ps})_+$.}
\label{Kac1}
\end{table}

\begin{table}
\begin{center}
\begin{tabular}{|c|c|c|c|c||c|}
\hline
   &  &  &  &  & $1$ \\
 \hline \hline
   &  &  &  &  & $\Omega L^3$ \\
\hline
   &  &  &  & $\Omega L$ & $\Omega^2 L^6$ \\
\hline
   &  &  &  & $\Omega^2 L^4$ & $ \Omega^3 L^9$ \\
\hline
   &  &  & $\Omega^2L^2$ & $\Omega^3 L^7$ & $ L^{12}$ \\
\hline
   &  &$\Omega^2$  & $\Omega^3 L^5$ & $L^{10}$ & $ \Omega L^{15}$ \\
\hline
   &  &$\Omega^3L^3$  & $L^8$ & $\Omega L^{13}$ & $\Omega^2 L^{18}$ \\
\hline
   &  $\Omega^3 L$ &$ L^6$  & $\Omega L^{11}$ & $\Omega^2 L^{16}$ & $\Omega^3 L^{21}$ \\
\hline
   &  $ L^4$ &$\Omega L^9$  & $\Omega^2 L^{14}$ & $\Omega^3 L^{19}$ & $ L^{24}$ \\
\hline
$L^2$  &  $\Omega L^7$ &$\Omega^2 L^{12}$  & $\Omega^3 L^{17}$ & $L^{22}$ & $\Omega L^{27}$ \\
\hline \hline
  $\Omega L^5$ &  $\Omega^2 L^{10}$ &$\Omega^3 L^{15}$  & $ L^{20}$ & $\Omega L^{25}$ & $\Omega^2L^{30}$ \\
\hline
\end{tabular}
\end{center}
\caption{\footnotesize Kac table for $(\hat p,\hat q; k)=(3,5;4)$. 
Thus $(p,q)=(6,10)$ and $d_{\hat q-\hat p}=2$. There are two copies of Kac table. 
The relation is $\mathcal O_{r,s}=(\bOmega^{r-s}\bL^{\hat qr-\hat ps})_+$, 
and the operators which do not correspond to the CFT one are obtained by just multiplying $\Omega$:
$\tilde {\mathcal O}_{r,s}=(\bOmega^{r-s+1}\bL^{\hat qr-\hat ps})_+$.}
\label{Kac2}
\end{table}

From the Kac table, one can easily see that the labeling $(\hat p,\hat q)$ of the minimal models 
can be identified as the order of the differential operator $(\bP,\bQ)$ of the multi-cut matrix-model, and 
the relation with the KP flows is the following:
\begin{align}
\mathcal O_{r,s} \quad \Leftrightarrow \quad \bB_{n=\hat qr-\hat ps}^{[r-s]} \qquad (\hat qr-\hat ps\geq 0).
\label{Operators}
\end{align}
This correspondence indicates the matching of the unphysical spectrum, 
\begin{align}
\mathcal O_{p,q-n} \quad \Leftrightarrow \quad \bB_{n\hat p}^{[n]}=(\bOmega \bL^{\hat p})^n, 
\qquad \mathcal O_{n,0} \quad \Leftrightarrow \quad \bB_{n\hat q}^{[n]}=(\bOmega \bL^{\hat q})^n, 
\end{align}
which means that the $(p,q)$ minimal fractional superstring theory corresponds to the following operators:
\begin{align}
\bP=(\bOmega\bL^{\hat p}),\qquad 
\bQ=\dfrac{(\hat q+\hat p)b_{\hat q+\hat p}^{[2]}}{\hat p}\, (\bOmega\bL^{\hat q})_+ +\cdots,
\end{align}
in the multi-cut matrix models. 
Note that the operator $\bQ$ does not start from $(\bOmega^{-1}\bL^{\hat q})$ like \eq{Qsym} 
which was derived in the assumption of the $\mathbb Z_k$ symmetry of the matrix model. 
This means that the background corresponding to this minimal model basically {\em breaks}
the original $\mathbb Z_k$ symmetry of the $k$-cut matrix model ($k\geq 3$), 
remaining at most $\mathbb Z_2$ symmetry of $(-1)^{r-s}$. 
This breaking symmetry property is actually an expected thing in the Liouville side, 
because the minimal-model correlators 
include the screening charges \eq{Screening} 
which belong to $R^{[2]}$ sector 
(and $R^{[k-2]}$ sector as the dual) and also breaks the $\mathbb Z_k$ symmetry, remaining 
at most $\mathbb Z_2$ symmetry.%
\footnote{In this sense, the two-dimensional fractional superstring theory (which might correspond to 
multi-cut matrix quantum mechanics) should preserve the $\mathbb Z_k$ symmetry. }
Note that the coincidence of this breaking nature is non-trivial 
since the origins of these phenomena in both sides are different. 

Finally we should note that the matrix model includes 
some operators which do not correspond to the operators of the 
conformal field theory in general. 
This is also observed in the case of $k=2$ for the odd model 
and such operators could be dropped by gauging a $\mathbb Z_2$ symmetry \cite{fi1}. 
Here we can drop these operators by gauging the following $\mathbb Z_k\times \mathbb Z_k$ symmetry, 
\begin{align}
(\bOmega, \bL)\quad \to \quad (\omega^{-(\hat qm-\hat pn)\hat k}\bOmega, \omega^{(m-n)\hat k}\bL) 
\qquad (m,n \in \mathbb Z),
\end{align}
and then the correspondence \eq{Operators} become a one-to-one mapping. 
Interestingly, this $\mathbb Z_k \times \mathbb Z_k$ symmetry is essentially 
the symmetry which preserves the form of the differential operators 
$(\bP,\bQ)=(\bOmega \bL^{\hat p},(\bOmega \bL^{\hat q})_+)$. 
We also note that before gauging this symmetry there are $d_{\hat q-\hat p}$ copies of the primary operators
of $(p,q)=(\hat k\hat p,\hat k\hat q)$ minimal fractional SCFT and they are related by the action of 
$\Omega^{l}$ ($l=1,2,\cdots, d_{\hat q-\hat p}-1$) on the gauge singlet operators. 

As a summary, the correspondence with respect to the spectrum can be phrased as 
{\em  $(p,q)=(\hat k\hat p,\hat k\hat q)$ minimal $k$-fractional superstring theory 
can be described with the $\mathbb Z_k$ breaking critical points of the $k$-cut two matrix model 
with $\bP=(\bOmega \bL^{\hat p})$ and $ \bQ=(\bOmega \bL^{\hat q})_+$.}
As we have seen, this correspondence requires several non-trivial matching of the spectrum structure. 
In the next subsection, we also consider coupling to the fractional super-Liouville system, 
and we will see that the string susceptibility of cosmological constant also coincides in both side. 

\subsection{Fractional super Liouville field theory}
Here we discuss the critical exponents, so coupling to fractional super-Liouville field theory. 
The original construction of Liouville theory \cite{Polyakov,DDK,DHK} 
always starts from the gauge fixing procedure 
of the (super-)diffeomorphism on worldsheet (super-)gravity. 
In the fractional super-Liouville case, we should also start from such a thing but we still do not know
how to define  ``fractional supergravity'' and even so-called ``fractional superfield formalism'' on 
``fractional superspace''. 

In the practical CFT calculation (for example \cite{fuku-hoso}), we do not need the 
other terms which include contact terms \cite{ContactTerms}.%
\footnote{For example, however, the minisuperspace formalism needs such terms \cite{NewHat}.}
So we assume that the action of fractional super Liouville field theory without contact terms
is given as 
\begin{align}
S_{Liou}=\frac{1}{2\pi k} \int d^2 z \Bigl( \del\phi\bar \del\phi 
+2\pi k\omega^{\frac{1}{2}}\, \rho\, \psi^L \tilde \psi^L\,  e^{\frac{2}{k}b\phi}\Bigr) 
+ S_{Z_k}(\psi^L,\tilde \psi^L), \label{LiouAct}
\end{align}
in conformal gauge and we only consider this action. Here we also use the $\alpha'=k$ convention. 
This action is the Liouville counterpart of the fractional supersymmetric sine-Gordon theory \cite{FSsineGordon}.
Here $\phi(z,\bar z)$ is the Liouville field, and $\psi^{L}(z)$ and $ \tilde \psi^{L}(\bar z)$ are 
the basic $\mathbb Z_k$ parafermion fields, with the statistics, 
$\tilde \psi^L(\bar z)\psi^L(z)=\omega\, \psi^L(z)\tilde \psi^L(\bar z)$. 
Let us call $\rho$ the cosmological constant. 
At the first sight, this choice of the cosmological constant term is strange 
because this operator belongs to ${\rm R}^{2}$ sector of the Liouville parafermion%
\footnote{See Appendix A. }
and NS sector (i.e. identity operator) of the matter parafermion.
However this choice is consistent with the string susceptability of the matrix model. 

From this action, we can obtain the energy-momentum tensor $T^L(z)$ and 
the fractional supercurrent $G^L(z)$ \cite{ALT,FSsineGordon} as 
\begin{align}
T^L(z)&=-\frac{1}{k}\bigl(\del \phi(z)\bigr)^2+\frac{Q_L}{k}\del^2\phi(z) + T_{Z_k}(z),\nn\\
G^L(z)&=\Bigl(\del \phi(z) -\frac{(k+2) Q_L}{4}\del  \Bigr) \epsilon(z) 
- \frac{k\tilde Q_L}{k+4}\eta(z),
\end{align}
with the background charges and the central charge, 
\begin{align}
Q_L=b_L+\frac{1}{b_L}, \qquad 
\tilde Q_L=b_L-\frac{1}{b_L}, \qquad 
\hat c_{L}\equiv \frac{k+2}{3k} c_L= 1+2\frac{(k+2)}{k^2}Q_L^2. 
\end{align}
If this system couples to matter CFT, the background charges $(Q_L,\tilde Q_L)$ should be related to 
those of matter CFT, $(\tilde Q,Q)$ in \eq{SuperChargeMatt}. It is natural to consider the following ansatz:
\begin{align}
Q_L\equiv Q=b+\frac{1}{b},\qquad \tilde Q_L\equiv \tilde Q=b-\frac{1}{b},\qquad b_L\equiv b. 
\label{CritiCharge}
\end{align}
This identification implies the critical central charge $\hat c_{crit}$ or 
the central charge of fractional superconformal ghosts, $\hat c_G$,
\begin{align}
\hat c_{crit}\equiv \hat c_L+\hat c_M=1+\Bigl(\frac{k+4}{k}\Bigr)^2 = -\hat c_G. \label{CritiDim}
\end{align}
This is actually nothing but the choice of the following critical central charge,
\begin{align}
-c_G=-\frac{3k}{k+2} \hat c_G= \frac{6k}{k+2} + \frac{24}{k}, \label{CritiCent}
\end{align}
which was found in \cite{ALT}.
So we just write $Q_L=Q$ and $\tilde Q_L=\tilde Q$ in the following discussion. 

The basic primary operators are 
$V_\alpha^{[\mu]}(z) \equiv \sigma_{\mu} (z):\!e^{\frac{2}{k}\alpha\phi (z)}\!:$ and 
their dual $(V_\alpha^{[\mu]} (z))^\dagger =V_{Q-\alpha}^{[k-\mu]}(z)$ with dimension: 
\begin{align}
\Delta(V_{\alpha}^{[\mu]}(z))=\Delta(V_{Q-\alpha}^{[k-\mu]}(z))= \Delta(\sigma_\mu)+\frac{1}{k}\alpha(Q-\alpha). 
\end{align}
Since the Seiberg bound \cite{SeibergNotes} in this case is also $\alpha \leq Q/2$, 
we choose the gravitational dressing \cite{DDK} of each primary operators 
to satisfy this bound. 
The KPZ-DDK exponents \cite{KPZ, DDK, DHK} of the correlators can be calculated as in the usual way, 
\begin{align}
\vev{\prod_{i=1}^N V_{\alpha_i}^{[\mu_i]}(z_i)}_\rho 
= \rho^{\ds \frac{Q}{2b}\chi-\sum_{i=1}^N\frac{\alpha_i}{b}} 
\vev{\prod_{i=1}^N V_{\alpha_i}^{[\mu_i]}(z_i)}_{\rho=1},
\end{align}
where $\chi$ is the Euler number $\chi=2-2h$ with the genus $h$. 

From this relation, 
we can obtain the scaling relation of the genus-zero partition function with respect to 
cosmological constant $\rho$: 
\begin{align}
\mathcal F_0(\rho) \sim \rho^{2-(1-b^{-2})} \quad \Leftrightarrow \quad 
\gamma_{str}^{(Liou)} = 1-b^{-2} = 1-\frac{\hat q}{\hat p} =\gamma_{str}^{(Mat)},
\label{MatchSUSC}
\end{align}
and this coincides with the matrix-model ``should-be'' cosmological-constant string susceptibility 
$\gamma_{str}^{(Mat)}$ of \eq{StrSusc}, 
with the identification of the pair of the coprime numbers $(\hat p,\hat q)$ 
as the order of the differential operators $\bP$ and $\bQ$. 
This is also consistent with our identification of the operator contents given in section 3.1.
Note that the consistent matching of string susceptibility depends on the choice of 
the cosmological constant term in the Liouville action \eq{LiouAct} and 
the critical central charge of \eq{CritiCharge}, \eq{CritiDim} and \eq{CritiCent}.

\subsection{The critical exponents and the on-shell vertex operators}

For further evidence of the correspondence, we need to consider the gravitational exponents of 
all the on-shell vertex operators. Actually this is not so easy now because the corresponding ghost system 
has not been known until now, and other quantization procedures and GSO projection 
also include some mysterious things \cite{AT1,ModelBuilding,CR,WSCFT,CCSS}. 
So we here give some proposal for the 
vertex operators and ghost primary fields which are consistent with the matrix models.%
\footnote{We should note that there is also another guess work for ghost system \cite{CR},
which is different from ours. }
So this should be justified by some other procedures in future investigation.



From the critical central charge \eq{CritiCent}, 
the ghost system is expected to have the following central charge%
\footnote{Here the system $Z_kG$ indicates the ``fractional super-partner'' 
of the $bc$ ghost system of $c_{bc}=-26$. Also note that this value has been also argued in 
\cite{ALT,ModelBuilding,CR}.}
\begin{align}
c_{Z_kG}\equiv c_{G}-c_{bc}=-2\,\frac{2(k-1)}{k+2} + 24\Bigl(1-\frac{1}{k}\Bigr),
\end{align}
and to have a set of the ``canonical''
ghost primary fields $\Xi^{(\lambda)}_\mu(z)$ 
($\lambda-\mu \in 2\mathbb Z$), 
with the identification,
\begin{align}
\Xi^{(\lambda)}_\mu(z)=\Xi^{(\lambda)}_{\mu+2nk}(z)
=\Xi^{(\lambda+nk)}_{\mu\pm n k}(z),
\end{align}
and the dimensions,
\begin{align}
\Delta(\Xi_{\mu}^{(\lambda)}(z))
=1-\frac{1}{k}-\Delta(\sigma_\lambda)-\Delta(\sigma_{\mu}) 
\equiv 1-a^{(\lambda)}_\mu
\qquad \bigl(\lambda-\mu \in 2\mathbb Z\bigr). \label{Intercept}
\end{align}
Here we call them as ``canonical'' because we can expect that 
there should be some generalization of ``picture'' in the superstring $\beta \gamma$ ghost system \cite{fms} 
and that they belong to the canonical picture primary operators.%
\footnote{
For example, the cosmological constant term in the action \eq{LiouAct} should belong to 0-picture. }
Also note that the value of $a^{(\lambda)}_\mu$ is 
the intercept $(L_0-a^{(\lambda)}_\mu)\ket{phys}=0$ of the 
fractional super-Virasoro constraints for the sector which couples to the ghost primary field 
$\Xi^{(\lambda)}_\mu(z)$, especially $a^{(\lambda+2)}_\lambda$ is the value for the massless 
${\rm R}^{[\lambda]}$ sector old covariant quantization discussed in \cite{ALT,NewKac}, 
which comes from the decoupling condition of the Lorentz-signature ghost contributions 
at the massless level. 

With these ghost primary fields, 
we can write down the tachyon-level vertex operator as the following $(1,1)$-primary operators: 
\begin{align}
\mathcal T_{r,s}(z,\bar z) = \Xi_{r-s}^{(r+s)} (z,\bar z) \,V_{r,-s}(z,\bar z)\, \mathcal O_{r,s}(z,\bar z) 
\qquad (\hat qr-\hat ps\geq 0). \label{OnShellTachyon}
\end{align}
Here $V_{n,m}(z)\equiv V_{\alpha_{n,m}}^{[n-m]}(z)$ is the special primary field with 
\begin{align}
\alpha_{n,m}=\frac{1}{2}\Bigl[(1-n)b^{-1}+(1-m)b\Bigr].  \label{LiouDegen}
\end{align}
The Seiberg bound indicates the restriction $\hat qn+\hat pm\geq 0$. 

If $mn>0$, then 
$V_{n,m}(z)$ corresponds to degenerate fields in Liouville theory and 
the null state operator appears at the level $N_{n,m}$, 
\begin{align}
N_{n,m}=\frac{nm}{k}+\Delta(\sigma_{[n+m]})-\Delta(\sigma_{[n-m]}),
\end{align}
and has the same dimension as $V_{-n,m}(z)$, that is, they are given as 
\begin{align}
\chi_{n,m}(z)=\Bigl(\mathcal Q_+^{(m)}\cdot V_{-n,m} (z)\Bigr)^\dagger 
= \Bigl(\mathcal Q_-^{(n)}\cdot V_{n,-m}(z)\Bigr)^\dagger \in \mathcal V_{n,m}.
\end{align}
It is worth to note the case of $b^2=\hat p/\hat q \in \mathbb Q_+$. 
In this case, the range indices $(p_L,q_L)$ of degenerate fields in Liouville theory appear, and 
they are the minimal integer pair of $(p_L,q_L)$ satisfying
\begin{align}
V_{n+lp_L,m-lq_L}(z)=V_{n,m}(z) \qquad (l\in \mathbb Z). 
\end{align}
The distinct thing of this case is that 
they are {\em different} from the conformal labeling $(p,q)$ of the matter CFT in general $(k\ge 3)$, 
and should be chosen as 
\begin{align}
(p_L,q_L)=(\hat k_L \hat p,\hat k_L \hat q),
\end{align}
with $k= \hat k_L\cdot d_{\hat q+\hat p}$. 
Here $d_{\hat q+\hat p}$ is the maximal common divisor 
among $k$ and $\hat q+\hat p$. 

There are several reason we choose the combination of the tachyon operators \eq{OnShellTachyon}. 
The first reason is that these operators have the following gravitational scaling dimension:
\begin{align}
\vev{\prod_{i=1}^l\mathcal T_{r_i,s_i}(z_i)} \sim \rho^{\sum_{i=1}^l\frac{n_i-(\hat q+\hat p)}{2\hat p}} 
\qquad (n_i\equiv \hat qr_i-\hat ps_i \geq 0),
\end{align}
which is consistent with the scaling dimension in the matrix model \eq{GravScale}, 
with the identification of \eq{Operators}: $\mathcal O_{r,s} \leftrightarrow \mathcal T_{r,s} 
\leftrightarrow \bB_{\hat qr-\hat ps}^{[r-s]}$. 
Another important check is about the cosmological constant vertex operator, 
\begin{align}
\mathcal T_{1,1}(z,\bar z)&=\Xi_{0}^{(2)}(z,\bar z)\, V_{1,-1}(z,\bar z)\, \mathcal O_{1,1}(z,\bar z)\nn\\
&= \Xi_0^{(2)}(z,\bar z)\, \sigma_{2}^L(z,\bar z) \,e^{\frac{2}{k} b\phi(z,\bar z)}.
\label{CosCan}
\end{align}
This belongs to ${\rm R}^{2}$ sector in the Liouville parafermion 
and has the same gravitational scaling dimension 
as that of the cosmological constant term in the action \eq{LiouAct},
\begin{align}
 \psi^L(z) \tilde \psi^L (\bar z) \, e^{\frac{2}{k} b\phi(z,\bar z)}, \label{CosPzAct}
\end{align}
which also belongs to the same ${\rm R}^2$ sector of the Liouville parafermion. 
From this we can argue the following thing:  

Here we recall that the picture changing operator in the superstring case 
is given as the combination of the ghost field $\xi(z)$ and supercurrent (i.e. gauge current) \cite{fms}: 
\begin{align}
X\equiv \int dz\, \xi(z) G(z), 
\end{align}
and that the fractional supercurrent $G(z)$ 
also belongs to NS sector whose 
action on other operators preserves the parafermion sector of the operators. 
This means that, even though the operators are expressed in the different pictures, 
the sector itself should {\em not} be changed. 
In this sense, our choice of the above cosmological constant operators in canonical picture 
(\eq{CosCan} and \eq{OnShellTachyon})
is consistent with the operator \eq{CosPzAct} in the action which can be related by some appropriate 
picture changing procedure. 

The other reason is related to 
the decoupling conditions for the Lorentz-ghost contribution 
discussed in \cite{ALT,NewKac}. 
The argument given in \cite{ALT,NewKac} is essentially the consideration of 
$(p,q)=(2,k+2)$ pure fractional supergravity system in our terminology,
and can be phrased as follows: 
Which dimension of the ghost primaries (i.e. the intercept $a_\lambda$) can enhance the null structure 
of the spectrum. The solution \cite{ALT,NewKac} is 
\begin{align}
\chi_\lambda^{(a)}(z )&=\Xi_{\lambda}(z )\Bigl[
\mathcal Q_{+}^{(\lambda+1+aq_L)}\cdot V_{-1-ap_L,\lambda+1+aq_L}(z)\Bigr]^\dagger 
\qquad (a=0,1,\cdots) \nn\\
\tilde \chi_\lambda^{(a)}(z )&=\Xi_{\lambda}(z )\Bigl[
\mathcal Q_{-}^{(aq_L-1)}\cdot V_{-1+ap_L,\lambda+1-aq_L}(z)\Bigr]^\dagger 
\qquad (a=1,2,\cdots),
\end{align}
with $\Delta(\Xi_\lambda)=a^{(\lambda+2)}_\lambda$ of \eq{Intercept}. 
That is, they are all $(1,1)$-primary operators and the Liouville primaries are null fields. 
Interestingly the matrix model implies that the set of ghost primary fields 
$\{\Xi_\lambda(z)\}_{\lambda=0}^{k}$ is not enough if we turn on the matter theory, 
and the generalization should be given as \eq{OnShellTachyon}.

With the choice of the primary fields \eq{OnShellTachyon}, 
there are the following enhancements of null structure:
\begin{align}
\chi_{r,s}^{(a)}(z)&=\Xi_{r-s}^{(r+s)}(z)
\Bigl[\mathcal Q^{(r+ap_L)}_- \cdot 
V_{r+ap_L,-(s+aq_L)}(z)
\Bigr]^\dagger \mathcal O_{r,s}(z),\nn\\
\tilde \chi_{r,s}^{(\tilde a)}(z)&=\Xi_{r-s}^{(r+s)}(z)
\Bigl[\mathcal Q^{((\tilde a+1)q_L-s)}_+ \cdot 
V_{-(p_L-r)-\tilde a p_L,(q_L-s)+\tilde aq_L}(z)
\Bigr]^\dagger \mathcal O_{r,s}(z),
\end{align}
with $a,\tilde a=0,1,\cdots$. That is, they are all $(1,1)$-primary fields of the following level:
\begin{align}
N_{r,s;L}^{(a)} =& 
\frac{1}{k}(r+a\hat k_L \hat p)(s+a \hat k_L \hat q)
+\Delta(\sigma_{r+s})-\Delta(\sigma_{r-s-a\hat k_L(\hat q-\hat p)}),\nn\\
\tilde N_{r,s;L}^{(\tilde a+1)} =& 
\frac{1}{k}(r-(\tilde a+1)\hat k_L \hat p)(s-(\tilde a+1) \hat k_L \hat q)
+\Delta(\sigma_{r+s})-\Delta(\sigma_{r-s+(\tilde a+1)\hat k_L(\hat q-\hat p)}).
\end{align}
We also note that the degenerate operator $\mathcal O_{r,s}$ has null states at the level 
\begin{align}
N_{r,s;M}^{(a_M)} =& 
\frac{1}{k}(r+a_M\hat k \hat p)(s+a_M \hat k \hat q)
+\Delta(\sigma_{r+s+a_M\hat k(\hat p+\hat q)})-\Delta(\sigma_{r-s}),\nn\\
\tilde N_{r,s;M}^{(\tilde a_M+1)} =& 
\frac{1}{k}(r-(\tilde a_M+1)\hat k \hat p)(s-(\tilde a_M+1) \hat k \hat q)
+\Delta(\sigma_{r+s-(\tilde a_M+1)\hat k(\hat q+\hat p)})-\Delta(\sigma_{r-s}),
\end{align}
and they turn out to be the same with the relations:
\begin{align}
N^{(t\hat k)}_{r,s;L}= N^{(t\hat k_L)}_{r,s;M},\qquad 
\tilde N^{(t\hat k)}_{r,s;L}=\tilde N^{(t\hat k_L)}_{r,s;M}, \qquad (t=1,2,\cdots). 
\end{align}
Although this enhancement is not the same kind argued in \cite{ALT,NewKac}, 
this should be related to the existence of the discrete states at the higher level 
\cite{LianZuckerman}. 

\section{Summary and discussion}

In this paper, we have pointed out that the non-critical $k$-fractional superstring theory can be described by
the $k$-cut matrix models. After we showed that multi-cut two-matrix model has 
the natural multi-cut matrix integral representation, we compared the operator contents 
of the matrix model and $(\hat p,\hat q)$ minimal $k$-fractional superconformal field theory. 
We then found that $(\hat p,\hat q)$ minimal $k$-fractional superstring theory corresponds to 
the critical point of $k$-cut matrix model with the differential operators 
$(\bP,\bQ)=(\bOmega \del^{\hat p}+\cdots, \bOmega \del^{\hat q}+\cdots)$. 
Several consistency checks are in order: 
\begin{itemize}
\item From the CFT point of view, the $\mathbb Z_k$ RR symmetry of the matter sector is broken 
by the screening charge and at most $\mathbb Z_2$ symmetry remains. We also observed 
this property in the matrix model side. That is, the critical point itself breaks the $\mathbb Z_k$ symmetry 
of the $k$-cut matrix models to at most $\mathbb Z_2$ symmetry. 
\item  Although the matrix model includes much more operators 
than $(\hat p,\hat q)$ minimal fractional superstrings, there is a 
$\mathbb Z_k\times \mathbb Z_k$ symmetry in the matrix model 
which can give the selection rule to correctly assign the operator contents of the CFT side. 
\item The definition of the cosmological constant operator is consistent with the matrix model, 
especially the string susceptibility with respect to the cosmological constant coincides on both sides. 
\end{itemize}
From this coincidence, we can again make sure 
that the critical central charge (or ghost central charge) \eq{CritiCent} argued in \cite{ALT} 
should be the correct value. 
At least from these primary fields \eq{Intercept}, 
the corresponding ghost system is like a ``ghost parafermion system'' which might be 
the wrong-statistics field theory of chiral (or ``Weyl'') parafermion.%
\footnote{
The ``wrong statistics'' means that the statistics is different from the canonical one only by sign $(-1)$.  
Also note that the usual parafermion system can be interpreted as ``Majorana'' parafermion 
because $k=2$ gives Majorana fermion. }
 It should be important to identify or 
concretely to construct the corresponding ghost CFT system. 

Although there are several non-trivial checks in this paper, 
the coincidence of the gravitational exponents of ``all'' the vertex operators 
is not totally accomplished. In this sense, 
it should be fair to say that this correspondence is still at the level of conjecture. 
So other relevant checks (for example, the coincidence of the D-brane amplitudes in both sides) 
should be important.  In particular, disk amplitudes \cite{SeSh} is interesting 
because this even does not require any knowledge of the ghost system. 
The interesting question is how the $\mathbb Z_k$ breaking nature does affect the $\mathbb Z_k$ 
charged FZZT brane and its algebraic curve.  It is also interesting to compare 
the annulus amplitudes because annulus amplitudes are sensitive to the RR charges \cite{Okuyama,Irie}.

Since the $\mathbb Z_k$-symmetry breaking critical points correspond to 
minimal $k$-fractional superstring theory, it is quite possible that there is some other $\mathbb Z_k$ 
symmetric minimal string theory which corresponds to the $\mathbb Z_k$ symmetric critical points 
of the $k$-cut two-matrix model. If there exists such a theory, 
the algebraic curve should show the $k$-cut geometry. 
Finding such a model should be also an interesting problem. 

The important future work in the matrix-model direction is direct evaluation of the critical points, 
especially how we fix the ``hermiticity'' of the KP flow parameters $t^{[\mu]}_n$. In the two-cut case, 
this was identified in \cite{HMPN}, that is, the odd potential should be pure imaginary. 
The multi-cut two-matrix model should also have such a selection of the potential and such a consideration 
can be only obtained from direct evaluation of the critical points. 

In this paper, we only focus on minimal fractional superstrings which can be described by the two matrix model. 
An interesting direction is to give an answer to the question 
what should correspond to multi-cut one-matrix model, 
which was originally given by \cite{MultiCut}, 
and also how we can construct the multi-cut matrix quantum mechanics 
which should describe the two-dimensional fractional superstring theory.

\vspace{1cm}
\noindent
{\bf \large Acknowledgment}  
\noindent
The author would like to express his appreciation to 
Masafumi Fukuma, Shigenori Seki and Yoshinori Matsuo 
for the fascinating collaborations and sharing insights into this field, 
and Jun Nishimura for suggesting of giving the talk at the KEK string workshop 2008 
which gives the starting point of this work, 
and Pei-Ming Ho and Kazuyuki Furuuchi for kind encouragements, 
and Yasuaki Hikida, Yoske Sumitomo, Shoichi Kawamoto, Hiroshi Isono, Darren Shih and 
Tomohisa Takimi for useful discussion, 
and Hikaru Kawai, Tamiaki Yoneya, Satoshi Iso, Tadashi Takayanagi, Jen-Chi Lee, Kazutomu Shiokawa 
and Yutaka Matsuo 
for useful comments, 
and David Shih for showing interests in this work, 
and Wen-Yu Wen, Dan Tomino and Hirofumi Mineo for various help in living in Taiwan, and 
finally Chuan-Tsung Chan for holding nice mini-school and study group. 
The author also would like to say thank you to the people in the KEK theory group and Taiwan string focus group.
This work is partially supported by Taiwan National Science Council under Grant No.~97-2119-M-002-001.

\appendix

\section{The parafermion and the $\mathbb Z_k$ spin-structure}
Here we summarize the basic ingredients and terminology for the parafermion CFT $Z_k$ \cite{ZF1} 
which we use in the section 3, especially focusing on how they can be the extension 
of the usual fermion. This terminology should be useful for this kind of ``parafermionic'' string theory. 
The conformal field theory of parafermion is parametrized by integer $k=1,2,\cdots$ 
and has the central charge of $c_{Z_k}=2(k-1)/(k+2)$. 
Basically this system is nothing but the coset CFT of $Z_k=SU(2)_k/U(1)_k$ \cite{ZF1}. 
Here we only focus on the left-hand-side (i.e. chiral) part of the CFT, and 
detail discussions of the parafermion can be found in \cite{ZF1, GepnerQiu}. 

\subsection{Primary fields and 
the $\mathbb Z_k\times \tilde{\mathbb Z}_k$ parafermion charge}
The basic dynamical degree of freedom is the parafermion field $\psi(z)$ 
with their descendants $\{\psi_l(z)\}_{l=0}^{k-1}$ of the dimensions, 
\begin{align}
\Delta_l\equiv \Delta(\psi_l)= \frac{l(k-l)}{k},
\end{align}
with $ \psi_0(z)=1, \ \psi_1(z)=\psi(z),\  \psi_{k-l}(z)=\psi_l^\dagger(z)$ and 
$\psi_{l+nk}(z)=\psi_l(z)$ $(n\in \mathbb Z)$. 
They form the following 
closed OPE algebra, the parafermion algebra:
\begin{align}
\psi_{l_1}(z)\,\psi_{l_2}(0)\sim 
\dfrac{c_{l_1,l_2}}{z^{\Delta_{l_1}+\Delta_{l_2}-\Delta_{l_1+l_2}}}\bigl[ \psi_{l_1+l_2}(0)+\cdots \bigr]. 
\end{align}
The dots in the parenthesis indicate 
descendants fields in the sense of $W_k$-algebra \cite{FZ-FL}. 
So we often use the abbreviate notation:
\begin{align}
\bigl[\psi_{l_1}(z)\bigr]_W\times \bigl[\psi_{l_2}(z)\bigr]_W = \bigl[\psi_{l_1+l_2}(z)\bigr]_W, \label{PFalg}
\end{align}
to clarify the algebraic structure. 
The important thing is that this algebra preserves the following $\mathbb Z_k$ symmetry:
\begin{align}
\Gamma^n:\, \psi_l(z)\, \mapsto\, \omega^{nl} \psi_l(z), 
\qquad \Gamma \equiv \omega^{f_R},\qquad 
\omega\equiv e^{\frac{2\pi i}{k}}. \label{PFnumber}
\end{align}
Here we call $f_R$ the worldsheet parafermion number (of left-hand-side). 
The charge $\Gamma$ is the counterpart of the usual chirality operator and 
is called the (worldsheet) parafermion charge operator. 

Other basic primary fields are spin fields $\{\sigma_\lambda\}_{\lambda=0}^k$ of 
the dimension 
\begin{align}
\Delta(\sigma_\lambda)=\frac{\lambda(k-\lambda)}{2k(k+2)},
\qquad \sigma_\lambda^\dagger(z)=\sigma_{k-\lambda}(z).
\end{align}
They are the vacuum operators of the parafermion module, 
\begin{align}
\bigl[\sigma_\lambda(z)\bigr]_\psi 
\equiv \bigoplus_{\mu\in \mathbb Z/\mathbb Z_k} \bigl[\sigma^{(\lambda)}_\mu(z)\bigr]_W,\qquad 
\bigl[\psi_{l}(z)\bigr]_W\times \bigl[\sigma^{(\lambda)}_{\mu}(z)\bigr]_W 
= \bigl[\sigma^{(\lambda)}_{\mu-2l}(z)\bigr]_W,  \label{PFmodule}
\end{align}
with 
\begin{align}
\sigma_\lambda(z) \equiv \sigma^{(\lambda)}_\lambda(z), \qquad 
\sigma^{(\lambda)}_\mu(z)=\sigma^{(\lambda)}_{\mu+2nk}(z) 
= \sigma^{(\lambda+nk)}_{\mu \pm nk}(z)\quad  (n\in \mathbb Z), \label{PFspinfield}
\end{align} 
especially $\psi_l(z)=\sigma^{(0)}_{2l}(z)=\sigma^{(k)}_{2l-k}(z)$. 
The dimension is 
\begin{align}
\Delta(\sigma^{(\lambda)}_\mu(z))=\frac{\lambda(\lambda+2)}{4(k+2)} -\frac{\mu^2}{4k}. 
\end{align}
Among them, there is the disorder (Kramers-Wannier) dual field $\mu_\lambda(z)$ 
for each $\sigma_\lambda(z)$:
\begin{align}
\sigma_\lambda(z) \quad \leftrightarrow \quad \mu_\lambda(z) 
\equiv \sigma^{(k-\lambda)}_{\lambda-k}(z), 
\end{align}
with $\mu_\lambda^\dagger(z)=\mu_{k-\lambda}(z)$. 
The assignment of the chirality $\Gamma$ for each primary fields has two choices in general:
\begin{align}
\Gamma(\sigma^{(\lambda)}_\mu(z))= \omega^{\frac{\mu+\lambda}{2}} \sigma^{(\lambda)}_\mu(z),
\qquad 
\tilde \Gamma(\sigma^{(\lambda)}_\mu(z))= \omega^{\frac{\mu-\lambda}{2}} \sigma^{(\lambda)}_\mu(z),
\label{ZkZk}
\end{align}
especially $\Gamma(\sigma_\lambda)=\omega^{\lambda}=\tilde \Gamma(\mu_{k-\lambda})$ and 
$\tilde\Gamma(\sigma_\lambda)=1=\Gamma(\mu_{k-\lambda})$. 
In this sense, the parafermion basically has the $\mathbb Z_k\times \tilde{\mathbb Z}_k$ structure in its spectrum.

\subsection{The $\mathbb Z_k$ spin-structure}

Although the $\mathbb Z_k$ symmetry \eq{PFnumber} is the symmetry of the algebra \eq{PFalg} and 
the module \eq{PFmodule}, this is not preserved in the general OPE algebra.%
\footnote{This charge is the coulomb charge of the Wakimoto construction and screened in general. }
This $\mathbb Z_k$ degree of freedom however appears in the twisted boundary condition of parafermion fields: 
\begin{align}
\psi(e^{2\pi i} z) \, O_\mu(0) = \omega^\mu\, \psi(z)\, O_\mu(0), 
\qquad \Omega(O_\mu(z))\equiv \omega^\mu O_\mu(z), 
\qquad \Omega \equiv \omega^F
\end{align}
and this $\mathbb Z_k$ charge $\Omega$ is preserved in the OPE algebra. 
This is the generalization of the spacetime fermion number or the R-R charge of the usual superstrings. 
We also call $\Omega$ the $\mathbb Z_k$ R-R charge or specetime ``parafermion number'', 
and we say that the operator $O_\mu$ belongs to ${\rm R}^\mu$ sector, 
and especially ${\rm R}^0$ sector is called NS sector. 
These sectors corresponds to the cuts of the parafermion fields in Riemann surfaces. 
We can easily see that the operator $\sigma^{(\lambda)}_\mu(z)$ belongs to ${\rm R}^\mu$ sector. 
The special operators of NS sector, $\epsilon_i(z)\equiv \sigma^{(2i)}_0(z)$ 
($\epsilon_i^\dagger (z) \equiv \sigma^{k-2i}_k(z)$), are called energy operators and have the
dimension
\begin{align}
\Delta(\epsilon_i(z))=\frac{i(i+2)}{k+2}. 
\end{align}
Note that this field has no $\mathbb Z_k$ R-R charge $\Omega$ but 
has the $\mathbb Z_k\times \tilde{\mathbb Z}_k$ parafermion charge $\Gamma \times \tilde \Gamma$. 

We also note that it is convenient to use the Young diagram notation 
based on the $\mathbb Z_k\times \tilde {\mathbb Z}_k$ structure of \eq{ZkZk}:
\begin{align}
\sigma^{(\lambda)}_\mu (z) \quad \Leftrightarrow \quad 
f_R \Biggl\{ \Young[-2]{22111} \raisebox{+3.5mm}{ \mbox{$\big\} \tilde f_R $}} \quad  \equiv  
{}^t(f_R,\tilde f_R),
\end{align}
which is actually related to the $W_k$-labeling. 
In this notation, the conjugation of $\Gamma \leftrightarrow \tilde \Gamma$ expresses the Kramers-Wannier duality, 
and the other conjugation $(\Gamma,\tilde \Gamma) \to (\Gamma^\dagger,\tilde\Gamma^\dagger)$ 
means the dual field conjugation. The size of the diagram is nothing but the $\mathbb Z_k$ R-R charge: 
$\Omega=(\Gamma\times\tilde \Gamma)_{diag}$.

\end{document}